# Ultrafast Coherence Delocalization in Real Space Simulated by Polaritons

Bo Xiang[1], Zimo Yang[1], Yi-Zhuang You[2]*, Wei Xiong[1,3,4]*

[1]Materials Science and Engineering Program, UC San Diego, 92093

[2]Department of Physics, UC San Diego, 92093

[3]Department of Chemistry and Biochemistry, UC San Diego, 92093

[4]Department of Electrical and Computer Engineering, UC San Diego, 92093

**Abstract** We investigated coherence delocalization on a coupled-cavity molecular polariton platform in time, frequency, and spatial domains, enabled by ultrafast two-dimensional infrared hyperspectral imaging. Unidirectional coherence delocalization (coherence prepared in one cavity transfer to another cavity) was observed in frequency and real spaces. This directionality was enabled by dissipation of delocalized photon from high-energy to low-energy modes, described by Lindblad dynamics. Further experiments showed that when coherences were directly prepared across cavities (superpositions between polaritons from different cavities), only energetically nearby polaritons could form coherences that survived the long-range environmental fluctuation. Together with the Lindblad dynamics, this result implied that coherences delocalized through a one-step mechanism where photons transferred from one cavity to another, shedding lights to coherence evolution in natural and artificial quantum systems. This work also demonstrated a way of combining photon and molecular modes to simulate coherence dynamics.

Coherences, a result from the superpositions of wavefunctions, are key ingredients of quantum systems, such as artificial quantum simulation platforms[1], natural light-harvesting antennas[2–9], ballistic energy transfer[10,11] and charge transfer materials[12,13]. To mediate non-diffusive energy transfer or transfer phase sensitive information, coherence propagates among different sites that are composed by distinct quantum states[4,8,10,14,15]. In systems such as light-harvesting complex and molecules that exhibit ballistic energy transfer, pioneering ultrafast spectroscopic studies have shown that coherence and excited populations move among different quantum states by resolving their signals in the frequency domain[6,9,10]. However, it remains to be a challenge to spatially track coherences in these systems and understand how coherences evolve when transported or delocalized into multiple sites, because most coherence transfer or delocalization in molecular systems happens between atoms that are a few angstroms or nanometers away[4,8,10,14,15]

In this work, using a coupled dual-cavity polariton platform[16], we simulate coherence delocalization and time-resolve its spatial distribution. When molecular vibrational modes are strongly-coupled to the cavity modes, they form hybridized quasiparticles – molecular vibrational polaritons[16–35]. This coherence delocalization is made available in polaritons due to the evanescent wave of the photon modes and anharmonicity of the molecular modes[16]. By preparing polaritons into specific quantum coherences, we visualize that coherence delocalizes from high to low frequency cavity, but the reverse is unfavored. Such a unidirectional transport is explained by photon delocalization and dissipation, and theoretical modeling shows that non-Hermitian Lindbladian dynamics is necessary to account the observed dynamics. To further understand the coherence spatial transport, we attempt to generate coherence directly between two cavities, referred as inter-cavity coherence. Only inter-cavity coherence between polaritons with sufficiently small energy separation (~ 10 cm$^{-1}$) could be prepared; these composed by polaritons in different cavities with larger energy difference (c.a. 30 cm$^{-1}$) are not observed experimentally. These results indicate that the spatial environment fluctuation only does not destroy phase relations between states that are energetically close enough.

The spatially-resolved coherence delocalization are conducted in molecular vibrational polaritons in liquid phase and at room temperature. Molecular systems are regarded unfavorable choices for simulating quantum phenomena, as their fast decoherence dynamics can eliminate quantum operations. In polaritons, the photons provide a route to bypass the decoherence limit, e.g., the fast photon speed relative to decoherence dynamics enable the simulation of coherence transfer. Similar idea of taking advantage of the photon velocity and nonlinearity of molecular modes in polaritons could open to a new direction of applying molecular species for simulating certain complex quantum and classical phenomena at room temperature.

**Results.**

In order to simulate coherence delocalization, we used a pair of laser pulses to create coherence between polaritons in one cavity and then followed the coherence evolution in real space and frequency by the probe pulse (Figure 1a). The measured signals, referred as two-dimensional infrared (2D IR) spectra[16,18,19,28–32,36], carried coherence and population dynamics of the coupled-cavity polaritons, which were imaged by a home-built hyperspectral microscope[37–39].

The coupled cavity enabled coherence delocalization and nonlinear interactions[16]. The cavity was formed by a distributed-Bragg-reflector (DBR) with checkerboard patterns and another flat DBR, which created alternating cavities with different longitudinal length. In our experiment, the depth of cavity A and B were 12.5 and 12.7 μm, respectively and the lateral width of each cavity was 50 μm (Figure 1b). To enable nonlinearity, we injected 40 mM W(CO)$_6$ in hexane solution into the coupled cavity. As shown in our previous work[16,28–32], the asymmetric vibrational modes of the molecules at 1983 cm$^{-1}$ strong coupled to the cavity that contained the molecules, and weakly coupled to the neighbor cavities by evanescent waves. Therefore, four polaritons (noted as UP$_i$ and LP$_i$ for polaritons in cavity i, where i = A or B) were formed (Figure 1c).

The polaritons were imaged using hyperspectral IR imaging, which magnified the signal by 12 times in real space through a pair of CaF$_2$ lenses and projected it onto the entrance of the spectrograph slit. The slit allowed a vertical cut of the image enter the spectrograph that relayed a one-to-one ratio image vertically and dispersed the spectra of each point along the vertical slice horizontally. In this way, the horizontal axis of hyperspectral images reported the spectra and the vertical axis reflected the spatial distribution of the signals (see Methods and Figure.2a). The image confirmed that two polaritons were formed in each cavity and they were vertically displaced from each other, reflecting the spatial separation between cavity A and B. The polaritons also diffused into the neighboring modes, indicating delocalization (Figure 1d. See Supplementary Information S1 for details).

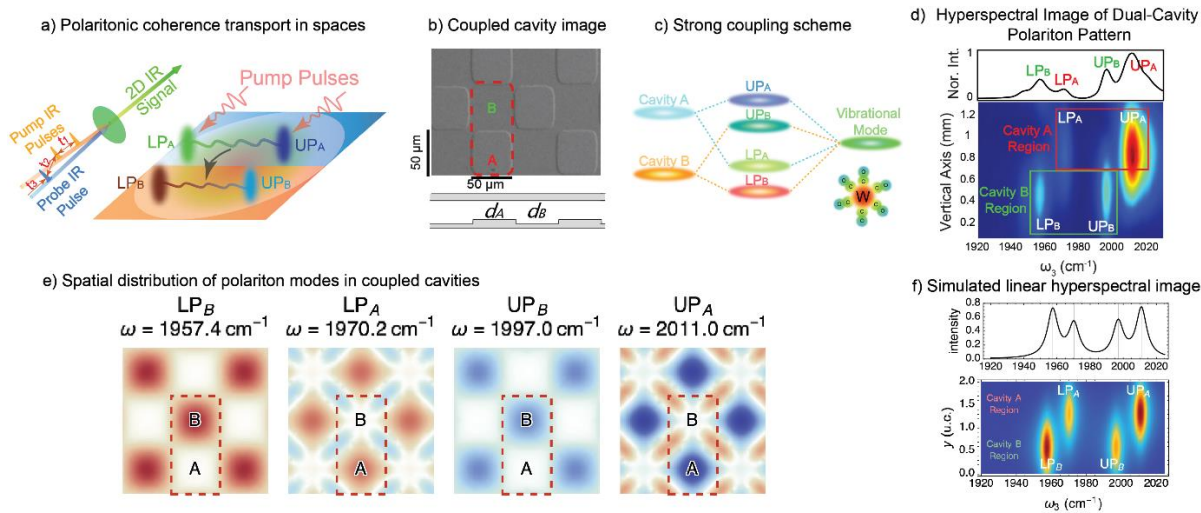

*Figure 1. **Overview of the idea of coherence delocalization in real spaces and dual-cavity polariton system.** (a) Schematic illustration of 2D IR pulse sequence and the general idea coherence delocalization between cavities. (b) SEM image of dual-cavity mirror (top view) along with diagrammatic showing the intersection. (c) Energy diagram showing the formation of two pairs of polaritons induced via strong coupling between the molecular vibrational mode of $W(CO)_6$ and cavity A and B modes respectively. (d) Hyperspectral image of coupled cavity polariton spatial distribution (vertical) and characteristic frequencies (horizontal). (e) Simulated spatial distribution of polariton modes in coupled cavities. (f) Simulated linear hyperspectral image with the corresponding transmission spectrum.*

To properly understand the polariton mode distribution, we simulated the cavity modes by the wave equation of electric fields where the coupled-cavity unit cell was modeled using parameters from experiments and assuming periodic boundary condition. Then we simulated the strong coupling between the cavity and molecular modes using Tavis-Cummings model[40] (see Materials and Methods and Supplemental Information S9). The simulated spectra (Figure 1f, top) showed four polariton peaks with the corresponding hyperspectral image (Figure 1f, bottom) revealing the polaritons from cavity A resided in the upper region whereas the ones from cavity B in the lower region. The model showed that polaritons mostly localized in their own cavities. However, the $UP_A$ and $LP_A$ states had distributions in cavity B (Figure 1e). As shown later, the propagation of $UP_A$ and $LP_A$ into cavity B were crucial for coherence delocalization.

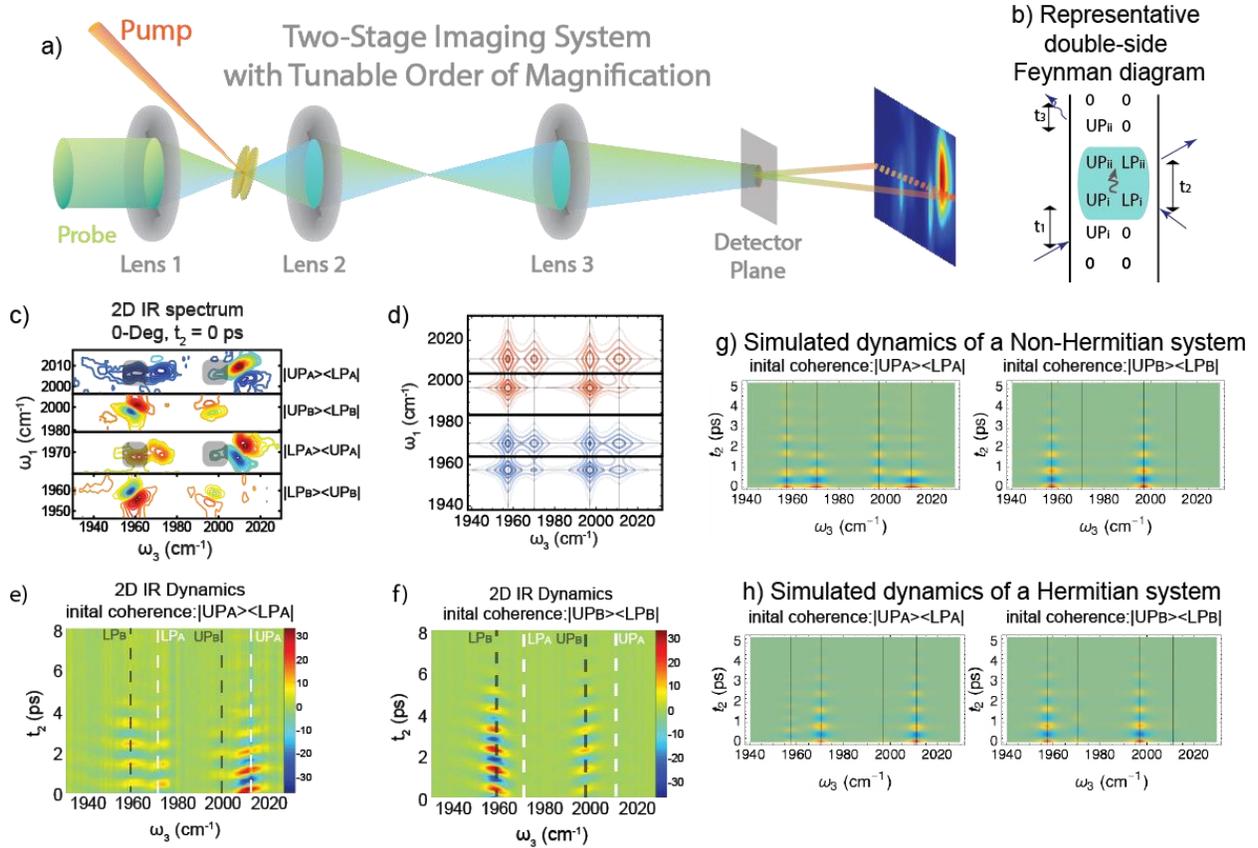

*Figure 2. **Intercavity coherence delocalization dynamics measured by 2D IR.** (a) Hyperspectral 2D IR-imaging setup. (b) A representative Feynman diagram showing the polariton coherence transfer. (c) Four 2D IR spectra with four different initial polariton coherence, from top to bottom: |UP$_A$><LP$_A$|, |UP$_B$><LP$_B$|, |LP$_A$><UP$_A$|, and |LP$_B$><UP$_B$|, where the 2D IR cross-peaks indicating the coherence transfer have been labeled in grey shaded area. (d) Simulated 2D IR spectrum of dual-cavity strong coupled system. 2D IR dynamics with initial coherence of (e) |UP$_A$><LP$_A$| and (f) |UP$_B$><LP$_B$|, the former shows clear coherence delocalization whereas the latter does not. Simulated 2D IR dynamics of initial coherence of |UP$_A$><LP$_A$| and |UP$_B$><LP$_B$| of (g) a non-Hermitian system and (h) a Hermitian system. Non-Hermitian Hamiltonian is necessary to reproduce experimental results.*

We next imaged the spatial coherent transport by an ultrafast hyperspectral IR imaging setup[37–39]. This instrument was done by a combination of 2D IR pulse sequence (Figure 1a) and imaging the signal by hyperspectral IR microscopy (Figure 2a). This setup allowed characterizing polaritons up to five-dimension (3D in frequency and 2D in space). In the following, we used it to spatially resolve 2D IR signals and coherence spatial distribution.

Because there were four polariton states in the system, many pathways could be populated when broadband pulses were used. To avoid unnecessary signals and focus on coherence delocalization only, we applied tailored pulse sequences to create specific coherences[30,41,42]. For example, we truncated the first two pulses in frequency domain so that it only created the initial coherence |UP$_i$><LP$_i$|, where i represented cavity A or B. Then, we scanned $t_2$ (the time delay between the second pump pulse and probe pulse, Figure 1a) to monitor the coherence evolution (Figure 2b). The representative spectra of all four initial coherences were listed in Figure 2c. It was noticeable that when |UP$_A$><LP$_A$| or |LP$_A$><UP$_A$| were created, there were also cross peaks at $\omega_3$= $\omega_{UP_B}$ and $\omega_{LP_B}$ (grey shaded areas in Fig. 2c), whereas such cross peaks were negligible when the other coherences were prepared.

We next examined the coherence dynamics. When the initial coherence was $|UP_A\rangle\langle LP_A|$, strong oscillating signals appeared at $\omega_3 = \omega_{UP_A}$ and $\omega_{LP_A}$, and $\omega_3 = \omega_{UP_B}$ and $\omega_{LP_B}$. All spectral features oscillated at ~40 cm$^{-1}$, corresponding to the Rabi frequency. Because we excluded the possibility of exciting polaritons in cavity B by the tail of the pump spectra (see supplemental information S5), the oscillating signals at $\omega_3 = \omega_{UP_B}$ and $\omega_{LP_B}$ suggested that coherence delocalized to cavity B upon being generated in cavity A. In contrast, when initial coherence $|UP_B\rangle\langle LP_B|$ was prepared, there were no noticeable delocalization to either $UP_A$ or $LP_A$. This unidirectional delocalization was robust, regardless of whether $|LP_A\rangle\langle UP_A|$ or $|UP_A\rangle\langle LP_A|$ was created, but neither $|LP_B\rangle\langle UP_B|$ nor $|UP_B\rangle\langle LP_B|$ could transfer to cavity A (see supplemental information S6). However, we found that when preparing and probing polariton coherences at higher in-plane momentum ($k_{||}$), such unidirectional delocalization became relaxed (additional experiments at higher $k_{||}$ are shown in supplemental information S7).

To examine whether polariton coherence indeed was delocalized from cavity A to B, we spatially-resolved the coherence dynamics of selective transitions (Figure 3). After firstly creating $|UP_A\rangle\langle LP_A|$, the oscillating signal at $\omega_3 = \omega_{UP_A}$ and $\omega_{LP_A}$ was localized in cavity A region (Fig.3a and b). However, at the same time, the signals at $\omega_3 = \omega_{UP_B}$ and $\omega_{LP_B}$ also oscillated at the same period and centered at cavity B. All coherence decayed within 3~4 ps, matching with the cavity lifetime. Therefore, these time-resolved hyperspectral images showed clear evidence that when coherence in cavity A was created, it delocalized into cavity B. Similar results were obtained when $|LP_A\rangle\langle UP_A|$ was created.

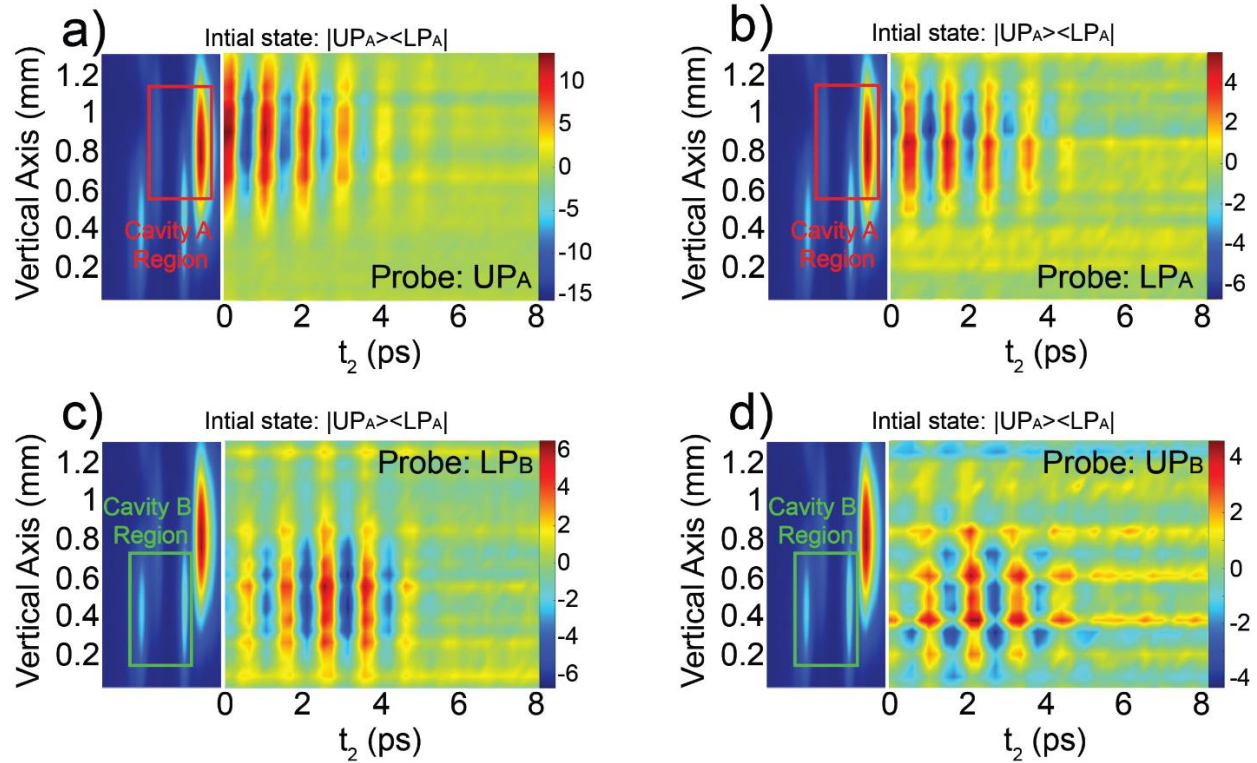

*Figure 3. **Spatially-resolved coherence dynamics obtained from 2D IR spectral cut and its comparison with linear transmission images (left).** Initial states are all $|UP_A\rangle\langle LP_A|$ and the probed states are (a) $UP_A$, (b) $LP_A$, (c) $LP_B$ and (d) $UP_B$.*

Several questions need to be answered about the unidirectional coherence delocalization, such as its origin. This unidirectional delocalization could be understood by the delocalization and dissipation of photon modes. The transfer of polariton coherence was driven by the propagation of photons from one cavity to another, supported by the associated ultrafast time scale. As photon transfer to another cavity, its energy should be conserved following the relations of $\boldsymbol{k}_A^2 + (n\pi/d_A)^2 = \omega^2/c^2 = \boldsymbol{k}_B^2 + (n\pi/d_B)^2$, where $c$ was the velocity of light in the cavity, $d_i$ and $k_i$ were the cavity thickness and in-plane momentum in cavity i. Because $d_A<d_B$, the lowest ($k_B \to 0$) photon mode in cavity B could not propagate to cavity A, since the propagation of photon required $k_A$ to be imaginary, implying attenuation of waves. Instead, when the lowest ($k_A \to 0$) photon mode in cavity A could hybridize with higher momentum modes in cavity B, as $k_B$ remained real and finite, supporting photon propagation from cavity A to B. Thus, the same principle that caused total internal reflections drove the unidirectional delocalization of photon.

However, the unidirectional photon delocalization alone was still insufficient to explain the transition of polariton in the frequency space, as the photon propagation was a Hermitian dynamic that preserved energy. Non-Hermitian energy dissipation dynamics[43,44] also played an important role. As photon propagated from cavity A to cavity B, its momentum increased, which allowed the photon to couple to the dark modes of molecular vibration in cavity B. The dark mode then decayed to the lowest momentum bright mode in cavity B by dissipating the vibration energy to the environment (i.e., solvents)[29,35]. In contrast, the reverse process was much harder, because the photon did not have enough energy to delocalize from cavity B to cavity A within the time scale of the process. In this way, the photons always decayed from the higher-frequency cavity to the lower-frequency one, but not the other way around. As a result, the polariton coherence was also transferred in the same unidirectional manner.

To justify our understanding, we further simulated the 2D IR spectra based on the Lindblad dynamic[45] of the density matrix with a non-Hermitian term[46] that described the photon decay from cavity A to cavity B. Although the phase twist in the experimental data were not reproduced in the simulation, (likely due to the inherit inhomogeneity of the polariton modes in experiments), the simulated spectrum in Figure 2d captured the salient features (such as peak position and intensities) of the experimental one. For example, the cross peaks that originated from coherence delocalization from cavity A to B was much more pronounced than the ones from the other direction. The simulated coherence dynamics in Figure 2g showed that the oscillation frequency of the |UP$_B$><LP$_B$| coherence was locked with that of |UP$_A$><LP$_A$|. However, if we turned off the non-Hermitian dissipation term in the Lindblad dynamics, the simulation in Figure 2h showed that the coherent transfer was no longer stable, and more importantly, the oscillation frequencies were no longer locked together (see dynamic cuts in figure S12). Because the energy was conserved in the absence of dissipation, such that each polariton mode would beat independently according to their detuning from the central frequency. The observation of the frequency locking behavior in the experiment strongly supported the importance of non-Hermitian dynamics in understanding the unidirectional coherence transfer.

We analyzed further the extend of the robustness of coherence across cavities. The intercavity distance was 50 μm, and solvent motion across this large distance was disordered and uncorrelated[47]. It was unclear how robust the phase could be against the fluctuation of solvents in real space and time. To gain insight to this question, we prepared coherences between polaritons in different cavities directly, e.g., |UP$_A$><UP$_B$| or |UP$_A$><LP$_B$| (Figure 4a and b). When either |UP$_A$><UP$_B$| or |LP$_A$><LP$_B$| was created, we observed a very slow oscillation at $\omega_2 \approx 10$ cm$^{-1}$ (corresponds to a period of 3~4 ps), which agreed with

the energy difference between the corresponding states (Figure 4c and d). The nonlinear oscillating signal was not only seen in the original excited polariton state, but also the other polariton from the same cavity. For instance, when $|UP_A><UP_B|$ were prepared, oscillating signals appeared at $UP_A$ and $UP_B$, and at $LP_A$ and $LP_B$ transitions. This was not surprising as the nonlinear interaction was strong between polaritons in the same cavity[32]. We noted the oscillating signal appeared at even higher frequency than $\omega_3 = \omega_{UPA}$ could be from polaritons of high order cavity modes, which was out of the scope of the current work. The existence of intercavity coherence such as $|UP_A><UP_B|$ and $|LP_A><LP_B|$ suggested that they could survive environmental variations between cavities.

A sharp contrast was when $|UP_A><LP_B|$ or $|LP_A><UP_B|$ was prepared, no nonlinear signal and coherence dynamics were observed (Figure 4e). This result indicated that such coherences did not survive the solvent variation across cavities. The energy separation between $UP_A$ and $LP_B$ was close to 50 cm$^{-1}$ and between $UP_B$ and $LP_A$ was about 30 cm$^{-1}$, significantly larger than the 10 cm$^{-1}$ when the intercavity coherences were observed. In general, high-frequency coherence was more vulnerable to fluctuations than low-frequency ones[48]. For example, it was easier to maintain coherence for mid-IR light than UV and x-ray. Similarly, this argument could explain why intercavity coherence of 10 cm$^{-1}$ existed while the one of higher frequency could not. Considering that neither $|UP_A><LP_B|$ nor $|UP_B><LP_A|$ could be created directly using laser pulses, the coherences would be destroyed if $|UP_A><LP_A|$ transferred to $|UP_B><LP_B|$ through these intermediate pathways (i.e. $|UP_A><LP_B|$ or $|UP_B><LP_A|$). This result suggested that multi-step coherence transfer across cavity was less likely. Instead, the coherence transfer was facilitated through a photon delocalization and dissipation process, where both the bra and ket states of the coherence finished transfer in one step.

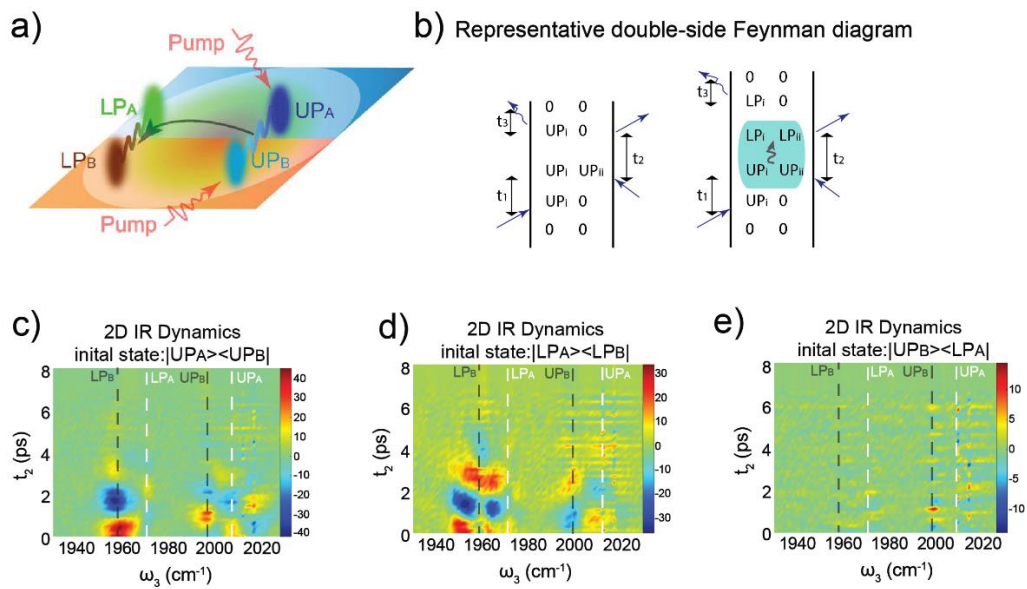

Figure 4. **Direct creation of intercavity coherence (superpositions between polaritons in difference cavities).** (a) illustration of $|UP_A><UP_B|$ formation as an example of intercavity coherence. (b) Feynman diagram showing the creation of intercavity coherence (left) and possible transfer pathways (right). 2D IR dynamics with initial states of, (c) $|UP_A><UP_B|$ and (d) $|LP_A><LP_B|$, (e) $|UP_B><LP_A|$, where $|UP_B><LP_A|$ shows nearly no signature of coherences, while the $|UP_A><UP_B|$ and $|LP_A><LP_B|$ show slow oscillation feature with a period of approximately 3 ps.

**Discussion.**

Taking advantages of photon delocalization and molecular nonlinearity, the coupled dual-cavity polariton was used to gain insights of coherence delocalization in real spaces, through a time-resolved hyperspectral IR microscope. The present study shed lights to coherence propagation in quantum systems. First, in a system which was at the bottom of its dispersion curve (i.e., in-plane momentum was zero), downhill coherence transfer was much favored than uphill. However, such preference was less strict at higher momentum space or when momentum conservation could be relaxed. Second, all coherences oscillated at the same frequency as the initial coherence, which could only be modeled by a non-Hermitian Linbladian dynamics, suggesting energy dissipation and photon delocalization both played critical roles in coherence transfer. Lastly, coherence could only be maintained between cavities when the energy difference between two quantum states was small enough to survive environmental variations in space. Thus, it was not likely that the coherence delocalized through any intermediate superposition states, with energy larger than 10 cm$^{-1}$. Instead, a one-step coherence transfer mechanism through single photon delocalization and dissipation was proposed. Such one-step mechanism suggested that in other coherence transfer systems, such as light-harvesting complex, the coherence delocalization could also depend on the spread of wavefunctions with the highest propagation speed.

The insights learned in this study could be applied when designing coherence transfer in new artificial quantum systems[1] or explaining coherence evolutions in natural quantum phenomena[11]. The present work also opened molecular polaritons as a new system for quantum simulation at ambient conditions. Molecular modes were often not considered for simulating coherent dynamics, as the stochastic processes in liquid phase at room temperature causes decoherence in the femto- to picosecond timescale. However, when strongly coupled with photons, whose velocity allow them to spread the coherence before it decays, the molecular vibrational modes can be used to understand certain ultrafast processes despite still being an open-dissipative. This advance of hybrid photonic-molecular modes can be further developed for simulating complex process complementing the capability of existing quantum simulation platforms, by developing advanced photonic structures, taking advantages of intrinsic molecular dynamics, such as energy transfer and isomerization[28,33,49,50], and multiple molecular chromophores.

**Materials and Methods**

**Preparation of Dual-Cavity Polariton System**

In order to generate two pairs of polaritons, a dual-cavity system has been developed with checkerboard patterns of 50-μm lateral dimension (see SI S1 for more details). The W(CO)$_6$ (Sigma-Aldrich)/coupled-cavity system is prepared by sandwiching W(CO)$_6$/hexane solution by a dual-cavity mirror and a flat dielectric CaF$_2$ mirror, separated by a 12.5 μm Teflon spacer. The W(CO)$_6$/hexane solution is nearly saturated concentration (40 mM).

**Theoretical modeling**

The cavity mode is simulated by solving the wave equation of electric field E(**r**) in the cavity

$$c^2\left(-\partial_x^2 - \partial_y^2 + \left(\frac{n\pi}{d(\mathbf{r})}\right)^2\right)E(\mathbf{r}) = \omega^2 E(\mathbf{r}),$$

where d(**r**) is the cavity thickness at position **r** = (x,y), c is the speed of light in the hexane solution inside the cavity, and n is the excitation mode quantum number in the perpendicular direction. Best fit to the observation data indicates that n = 4. We focus on the bright mode solutions of the wave equation, which correspond to modes with zero quasi-momentum over the cavity lattice and s-wave symmetry inside the cavity. We find two such modes at $\omega_A$ = 1998.2 cm$^{-1}$ (dominantly in cavity A) and $\omega_B$ = 1971.4 cm$^{-1}$ (dominantly in cavity B). Based on these two cavity photon modes, the polariton modes can be modeled by the Tavis-Cummings model described by the following Hamiltonian

$$H = \sum_{i=A,B} \omega_i a_i^\dagger a_i + \omega_0 \sigma_i^+ \sigma_i^- + g(a_i^\dagger \sigma_i^- + h.c.),$$

where $a_i$ denotes the photon annihilation operator of the $i$th cavity mode and $\sigma_i^\pm$ denotes the raising/lowering operator of the (collective) vibration mode that couples to the corresponding photon mode. Dark modes will be omitted in the model. The best fit to the experimental observation shows that the light-matter coupling strength is around $g$ = 18.7 cm$^{-1}$. Using this parameter, our model produces polariton modes at $\omega_{UPA}$ = 2011.0 cm$^{-1}$, $\omega_{UPB}$ = 1997.0 cm$^{-1}$, $\omega_{LPA}$ = 1970.2 cm$^{-1}$, and $\omega_{LPB}$ = 1957.4 cm$^{-1}$, matching the experimental observation nicely. We assume the pump/probe laser can be modeled by the perturbation

$$V = \sum_{i=A,B} \mu_i (a_i^\dagger + a_i),$$

where $\mu_i$ characterizes coupling strength between the laser mode and the cavity mode. To simulate the pump-probe dynamics, we consider the Lindblad dynamics of the polariton, under which the density matrix $\rho$ of the system evolves by

$$i\partial_t \rho = \mathcal{L}[\rho] = [H,\rho] + i\sum_m \left(F_m \rho F_m^\dagger - \frac{1}{2}\{F_m^\dagger F_m, \rho\}\right),$$

where the Lindblad operator $F_m$ enables us to describe the dissipation of the photon and vibration modes by $F_{1,2,3,4} = \sqrt{\Gamma_1} a_A, \sqrt{\Gamma_2} a_B, \sqrt{\Gamma_3} \sigma_A^-, \sqrt{\Gamma_4} \sigma_B^-$, as well as the non-Hermitian photon transfer from high-energy to low-energy modes $F_5 = \sqrt{\Gamma_5} a_B^\dagger a_A$. Starting from the initial vacuum state $\rho$ = |0><0|, the pump/probe laser acts by $\rho \to i [V, \rho]$, the time evolution is generated by the Liouvillian super-operator $\rho \to \exp(-i L t)[\rho]$, and finally, the emission amplitude is given by Tr $V \rho$. Within this formalism, we can simulate the 2D IR spectroscopy result and compare with experiment. It turns out that the $F_5$ non-Hermitian photon transfer process plays important role in understanding the coherence transfer between cavities. Additional theoretical details can be found in the Supplementary Information S9.